\documentclass{elsart}

\usepackage{natbib}

 \usepackage{epsfig}

\usepackage{amssymb}

\begin{document}

\begin{frontmatter}



\title{Status and perspectives of indirect and direct dark matter searches}


\author{Nicolao Fornengo}
\address{Department of Theoretical Physics, University of Torino, and INFN--Sezione di Torino,
via P. Giuria 1, I--10125 Torino, Italy \\ fornengo@to.infn.it - http://www.to.infn.it/$\,\tilde{~}fornengo$}

\begin{abstract}

In this review article the current status of particle dark matter is
addressed. We discuss the main theoretical extensions of the standard
model which allow to explain dark matter in terms of a (yet
undiscovered) elementary particle. We then discuss the theoretical
predictions for the searches of particle dark matter: direct detection
in low background underground experiments and indirect detection of
neutrinos, gamma-rays and antimatter with terrestrial and space-borne
detectors. Attention will be placed also on the discussion of the
uncertainties, mainly of astrophysical origin, which affect the
theoretical predictions. The constraints placed by these searches on
the extensions of the standard models will be briefly addressed.

\end{abstract}


\end{frontmatter}

\section{Introduction}

Large amounts of dark components have been clearly identified in our Universe,
on different scales and by different experimental means. The current view of 
modern Cosmology sees the Universe very close to being flat, with 30\% of its content
in the form of a cold dark matter (CDM) component, responsible for structure formation, while 
the remaining 70\% made of a very exotic dark energy component which causes its recent accelerated expansion. Baryons can account at most 4--5\% of the total content of
the Universe, much less than the CDM amount. This fact points toward a non--baryonic
origin of dark matter and this is a clear evidence that our understanding of the
elementary particle physics component of matter, beautifully described by the
Standard Model (SM), is incomplete. We need to extend the particle content in order to
accomodate (at least) one non--baryonic dark matter candidate, since the only DM
candidate in the SM is the neutrino which is unsuited to explain the bulk of DM since
it acts as hot dark matter and instead CDM is largely required to succesfully produce the
observed large scale structure of the Universe.

Supersymmetry offers a wonderful possibility, since the lighest supersymmetric particle
is stable, once R--parity is conserved, and it naturally posseses the properties of
a succesfull CDM candidate (neutrality and weak interactions) in many realization of suspersymmetry. The most successful and studied candidate is the neutralino, and 
I will concentrate on this particle in the following, where I will give a brief overview
of the strategies for neutralino dark matter searches and of some recent results. 
For a more detailed discussion and a more exhaustive list if references, I cross--refer
the reader to the quoted papers.

\section{Strategies for dark matter searches}

There are two basic ways to detect WIMP (Weakly Interacting Massive Particles) 
dark matter which is present in the halo of our Galaxy. The first method, direct detection, relies on the possibility to detect the recoil energy of the nuclei of a low--background detector as a consequence of their elastic scatting with a WIMP. The second method, indirect detection, exploits the possibility to detect products of the annihilation of DM particles, either
in the galactic halo or in celestial bodies (namely the Earth and the Sun) where WIMPs
may have been accumulated by gravitational capture. In this last case, the signal consists
of a flux of neutrinos emitted from the central regions of the body, and the typical
observable is a flux of upgoing muons produced by the charged--current conversion of the muon neutrino component of the signal. In the case of DM annihilation in the galactic halo,
there are more possibilities: the signal can consists of gamma--rays, neutrinos and
antimatter (positrons, antiprotons and antideuterons).

From the experimental side, the searches of DM signals involve many different techniques,
ranging from low--background underground detectors, to neutrino telescopes, antimatter
and gamma--rays detectors in space, to air-Cerenkov detectors.

\section{Direct detection}
\begin{figure}
\label{figure1}
\begin{center}
\includegraphics*[width=8cm]{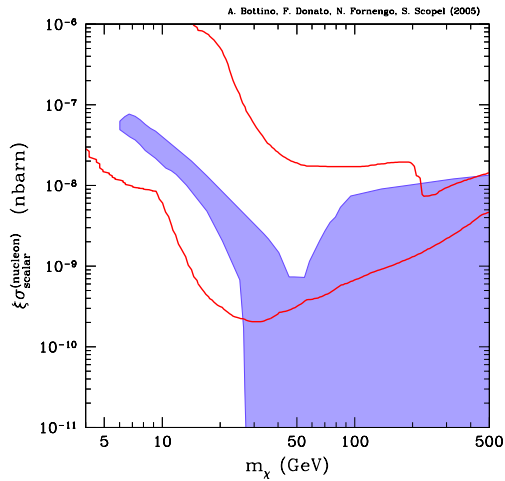}
\end{center}
\caption{Direct detection scattering cross section on a nucleon
vs. the WIMP mass. The solid lines show the allowed region by the
DAMA/NaI experiment compatible with the observed annual modulation
effect (\cite{dama}) and derived for a wide variation of galactic halo
models (\cite{dama}). The shaded area shows theoretical predictions for
neutralino dark matter in a low--energy supersymmetric standard
models. Configurations for masses below about 45 GeV refer to gaugino
non--universal models, while for higher masses gaugino--mass
universality is assumed (\cite{noidirect}).}
\end{figure}
\begin{figure}[t]
\label{figure2}
\begin{center}
\includegraphics*[width=8cm]{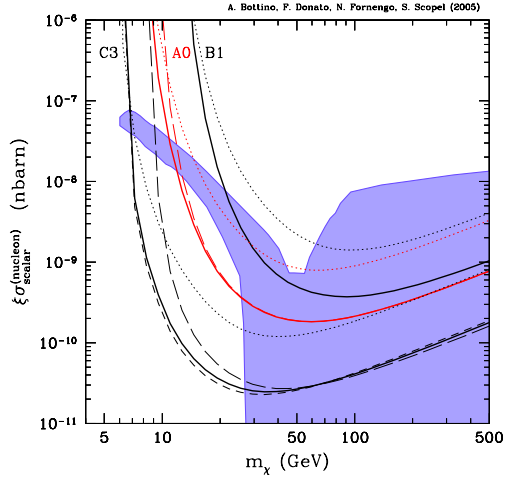}
\end{center}
\caption{The same as in Fig. 1, except that the solid lines denote
upper limits from the CDMS detector (\cite{cdms}) for some specific galactic halo
models, as calculated in \cite{noidirect}.}
\end{figure}
\begin{figure}[t]
\label{figure3}
\begin{center}
\includegraphics*[width=13cm]{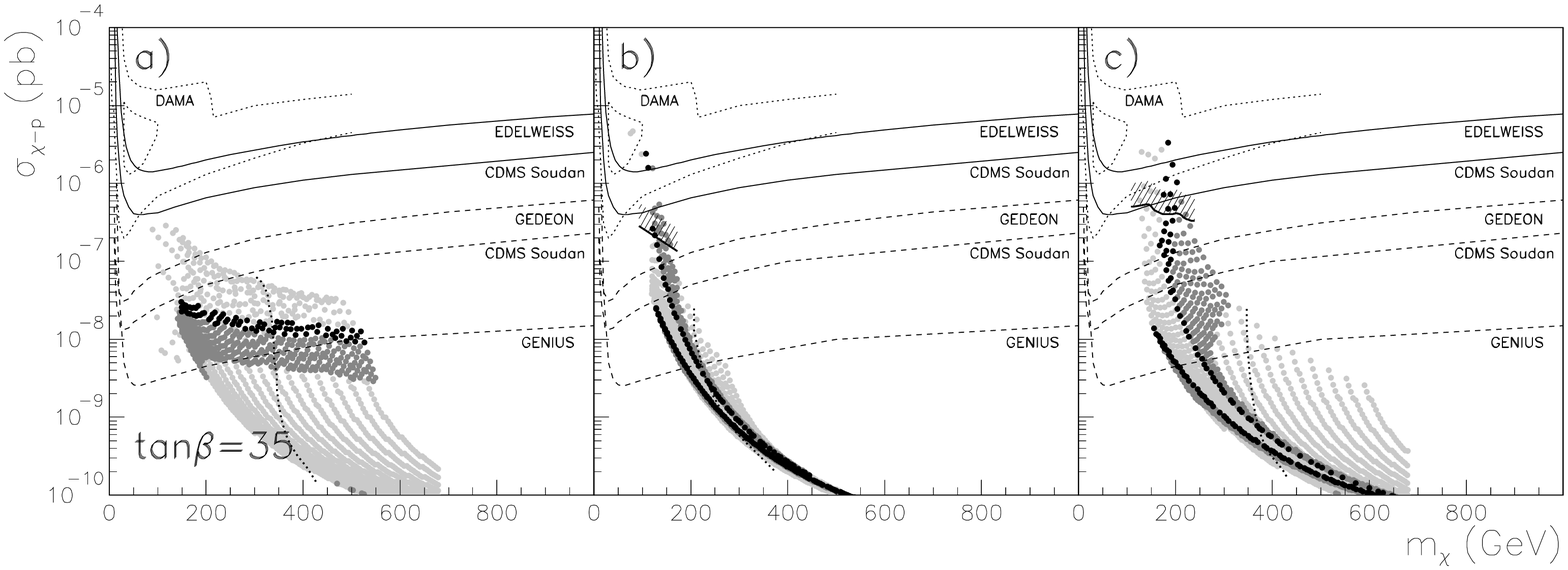}
\end{center}
\caption{Direct detection scattering cross section on a nucleon
vs. the WIMP mass. The theoretical predictions are here obtained in
non--minimal SUGRA, where non--minimality refers to the Higgs sector.
Darker points refer to configurations which match the cosmological
abundance for CDM. Solid lines refer to current experimental results
and future prospects (figure from \cite{munoz}).}
\end{figure}

From the particle physics point of view, direct detection relies on the
scattering cross section of the WIMP with the nucleon in the nuclei of the
detector. Experimental results have reported a positive indication of a signal
in terms of the annual modulation of the rate due to the Earth motion relative
to the WIMP wind: the DAMA/NaI Collaboration has a clear detection
of a temporal modulation with the expected amplitude, phase and period \citep{dama}.
When interpreted as due to dark matter scattering, the allowed region shown in Fig. 1
is obtained for the scattering cross section vs. the WIMP mass. Figure 1 refers to the
case of coherent WIMP--nucleus scattering. The same figure shows the comparison of the
DAMA/NaI annual modulation region with predictions for neutralinos obtained in two different supersymmetric models \citep{noidirect}. The part of the shaded area which refers to neutralino masses larger than about 50 GeV refers to a low--energy (electroweak scale) realization of the minimal supersymmtric standard model (MSSM). In this models a lower mass bound of about 50 GeV is obtained
from LEP searches. Once the gaugino--universality condition, usually assumed in these
models, is relaxed, the LEP bound loosens and a lower limit of about 6 GeV on the neutralino mass is obtained instead by Cosmology, requiring that neutralinos do not contribute to
the CDM content of the Universe in excess of the experimental upper bound \citep{noilow}. In Fig. 1, the configurations relative to these gaugino non--universal models are those relative to masses below 50 GeV. It is clearly seen that the direct detection cross section predictions
are sizeable and able to expain the DAMA/NaI result easily \citep{noidirect}.

Figure 2 show the same theoretical predictions confronted against upper limits obtained by the
CDMS detector \citep{cdms}. The upper limits are here re--calculated in order to show the
sizeable dependence of direct detection on the phase space properties of WIMPs in the galactic halo \citep{noidirect}.

Figure 3 is an example of theoretical prediction in a different supersymmetric models, namely
a minimal Supergravity (SUGRA) scheme with non--universality in the Higgs sector \citep{munoz}. In these type of models, especially in strict universal SUGRA, typically the predictions are lower than in the case of low--energy MSSM. Non--universalities are instrumental in obtaining larger detection rates.


\section{Antiproton signal}
\begin{figure}[t]
\label{figure4}
\begin{center}
\includegraphics*[width=8cm]{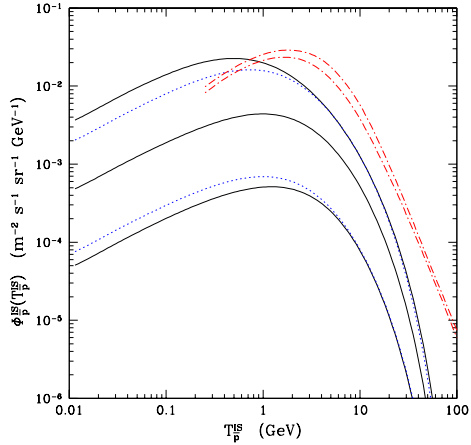}
\end{center}
\caption{Antiproton flux as a function of antiproton kinetic energy. The dot--dashed lines 
refer to the secondary production from cosmic ray spallation, i.e. the background for dark matter
studies. The solid lines show a prediction of a dark matter antiproton signal, for a WIMP of
100 GeV mass. The central solid line refers to the best choice of astrophysical parameters
for galactic cosmic ray propagation. The upper and lower solid (dotted) lines show the uncertainty band
at $4\sigma$ ($2\sigma$) C.L. \citep{noipbarsalati}.}
\end{figure}
\begin{figure}[t]
\label{figure5}
\begin{center}
\includegraphics*[width=8cm]{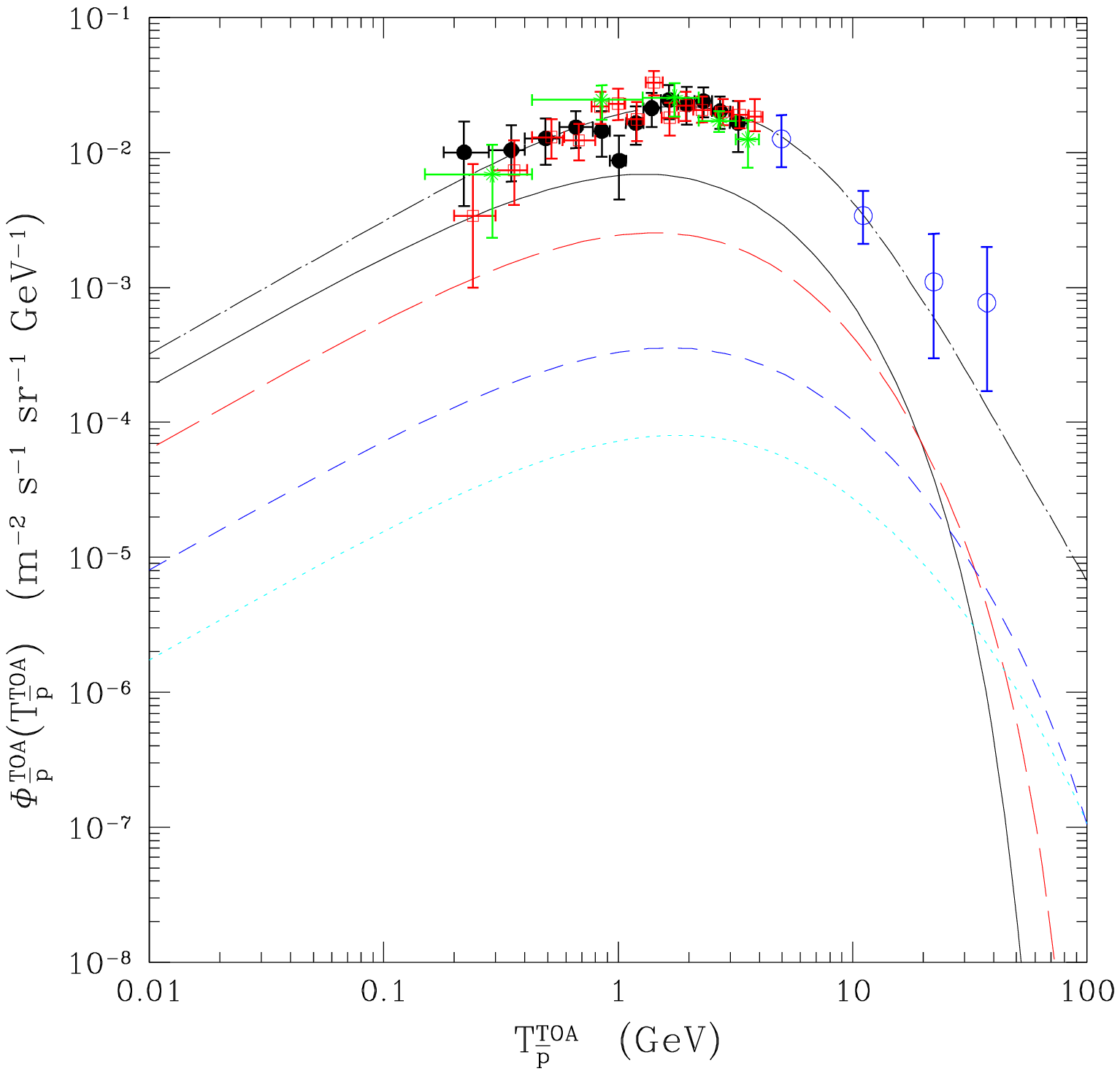}
\end{center}
\caption{Comparison of the theoretical predictions for the antiproton flux and experimental data \citep{noipbarsalati}.
The full circles/open squares/dots/open circles show the data from 
BESS95--97\citep{bess95}/BESS-98\citep{bess98}/AMS\citep{ams}/Caprice\citep{caprice} experiments, respectively.
The upper dashed line shows the background prediction. The other lines show predictions from dark matter
annihilation, for different WIMP masses \citep{noipbarsalati}.}
\end{figure}
\begin{figure}[t]
\label{figure6}
\begin{center}
\includegraphics*[width=8cm]{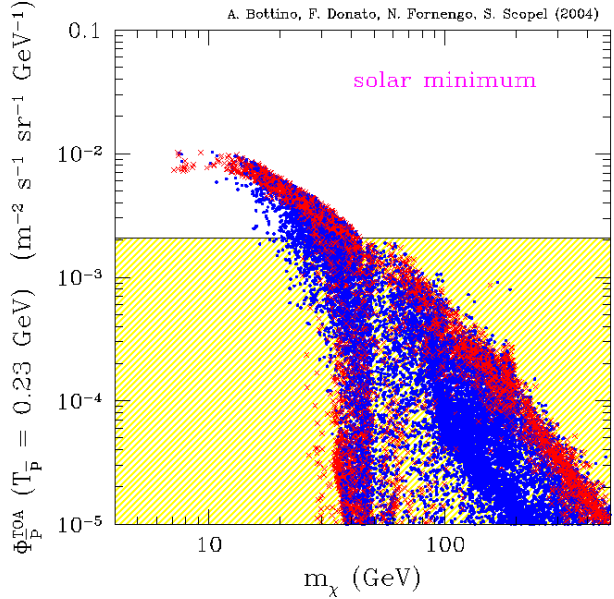}
\end{center}
\caption{Antiproton flux at kinetic energy $T=0.23$ GeV as a function of the neutralino mass, calculated
in the low--energy minimal supersymmetric standard model. For masses lighter than 50 GeV, the model
allows for non--universality in the gaugino sector; for higher masses, standatd GUT universality holds.
The shaded area denotes the amount of exotic
antiprotons which can be accomodated at this energy without conflicting with the existing data and background
calculations. The astrophysical propagation parameters are set at their best--fit values \citep{noiindirect}.}
\end{figure}
\begin{figure}[t]
\label{figure8}
\begin{center}
\includegraphics*[width=8.5cm]{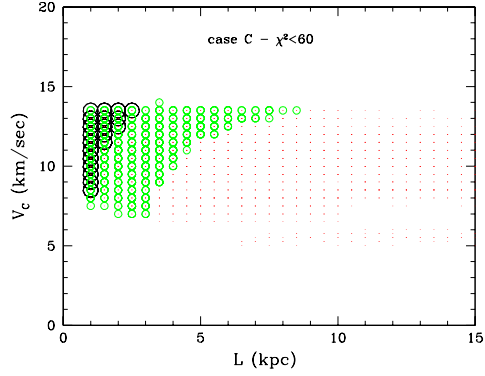}
\end{center}
\caption{Regions in the astrophysical--propagation--parameters space where the fit to the
experimental antiproton data of a background+signal--component is statistically acceptable.
In this case the plane convective velocity $v_c$ vs. the height $L$ of the diffusve region is shown.
Big circles, small circles and dots refer to neutralino masses of 10, 20, 30 GeV, respectively.
\citep{noipbar}.}
\end{figure}
\begin{figure}[t]
\label{figure7}
\begin{center}
\includegraphics*[width=8cm]{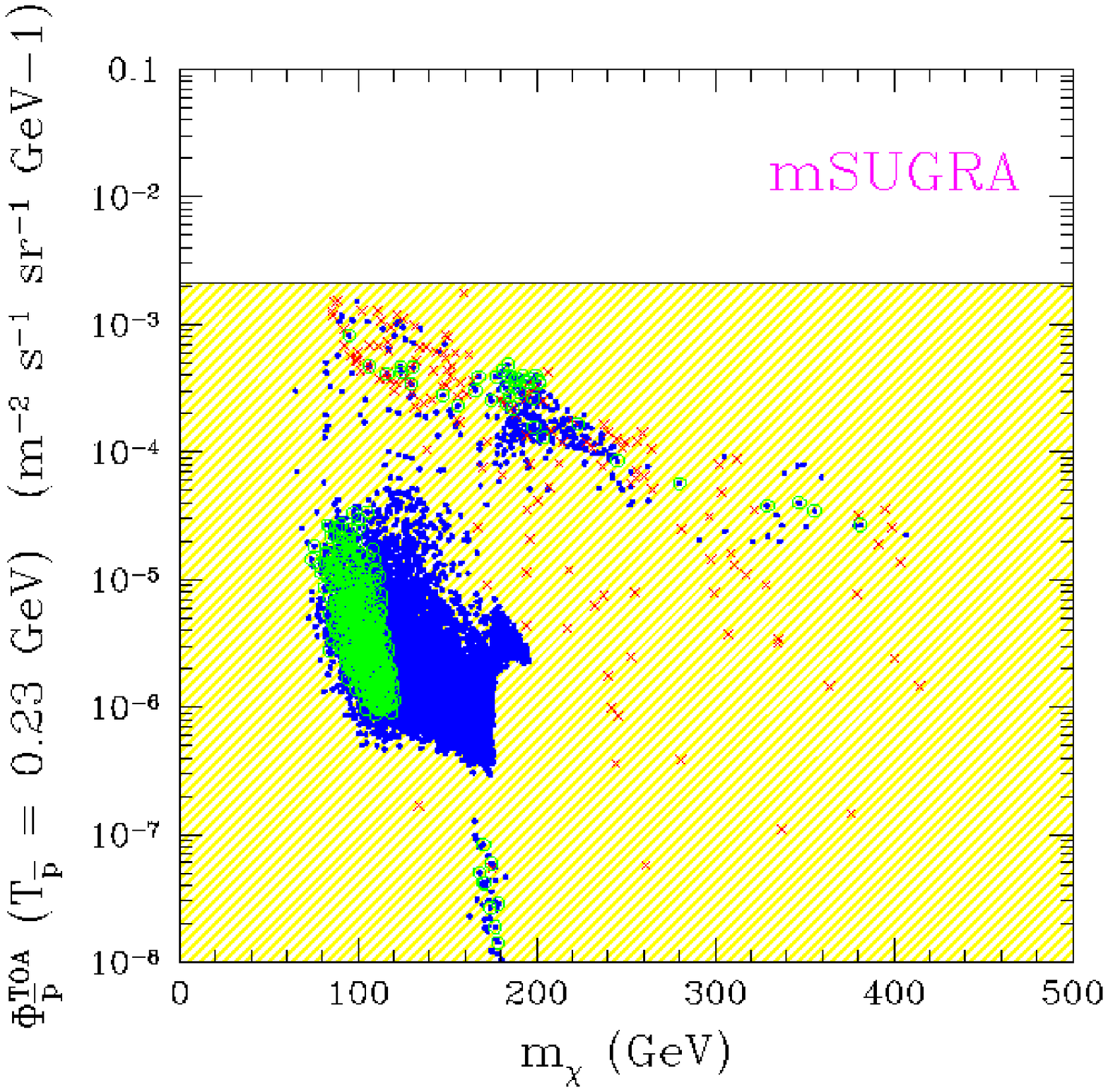}
\end{center}
\caption{The same as in Fig. 6, calculated in a minimal (universal) SUGRA scheme \citep{noipbarsalati}.}
\end{figure}

Antiprotons may be produced by WIMP annihilation in the galactic halo. Once they are produced, they
suffer a complicated mechanism of propagations inside the galactic diffusive halo, process which involves
both diffusion and energy losses. Once they reach the boundary of the heliosphere, antiprotons further suffer
propagation against the solar wind, which changes the low--energy part of the antiproton spectrum and leads to
the phenomenon of solar modulation, related to the 11 year solar cycle. All these diffusive and energy redistribution
processes are properly taken into account by solving the diffusion equations in the specific medium. In the case
of galactic propagation, a detailed study of the propagation processes for the DM antiproton signal has shown that
the theoretical predictions suffer of large uncertainties in the low energy tail, uncertainties which are related to
the poor knowledge of the propagation astrophysical parameters and which have been
quantified to an order of magnitude above and one below the median estimate \citep{noipbarsalati}.

A standard component of cosmic antiprotons is also produced by standard cosmic ray processes: this component
is the background for this type of dark matter studies. Contrary to the case of the signal, the standard
antiproton component suffers from much smaller uncertainties (see \cite{noipbarsalati} and references therein),
of the order of 20--30\%. The problem of antiproton searches is therefore to disentangle a signal in the low
energy tail, signal which suffers at the moment from large uncertainties, from a much more under control background.
The current experimental data on the antiproton flux vs. the antiproton kinetic energy are plotted in Fig. 5
together with the theoretical prediction for the background and some examples of predicted signals from
neutralino annihilation. Fig. 5 shows two key features: the first is that the theoretical estimate of the
background agrees well with the experimental data, which means that we cannot accomodate a large exotic
component; the second is that the low energy tail of the spectrum, which is where the signal could
mostly reveal itself, has very similar shape as compared to the background, a feature which prevents antiprotons
from having a clear signature. A flux of antiprotons from dark matter annihilation coming out from the background
could show up at energies larger that a few tens of GeV (a region where no data are currently available and
which will be covered in the future by Pamela and AMS): nevertheless this possibility requires somewhow large
dark matter overdensities, since in this case the WIMP must have a large mass and the WIMP number density scales rapidly with increasing DM mass (the signal scales even faster, with the number density squared).

Concentrating on the low energy tail of the antiproton flux, we may therefore derive constraints on the exotic
DM production and use this bound to constrain suspersymmetric models. Antiprotons represent the best indirect
signals for constraining dark matter searches, even though we must remember the large astrophysical uncertainty
which they suffer. Fig. 6 shows an example for neutralinos in the low--energy MSSM with (for masse above 50 GeV)
and without (for masses below 50 GeV) gaugino universality. The plot is shown for the best--fit value of the
astrophysical propagation parameters: uncertainty of one order of magnitude above and below the plotted points
must be taken into account. Even in this unfortunate situation we see that antiprotons searches can offer a quite
strong constraint on low mass neutralinos. This is better shown in Fig. 7, where we show the regions in the
astrophysical parameter space which are compatible with antiprotons produced by low--mass neutralinos. We see that
the predicted signals are compatible with observations only is a very limited and correlated sector of the
astrophysical parameter space. This situations is expected to improve significantly in the near future: therefore
we could be able to set clear constraints (or detect a signal), especially for light mass neutralinos.

Fig. 8 shows an example of theoretical calculations in a different supersymmetric model, namely a minimal
SUGRA scheme with full universalities. In this case the predicted signals are typically lower than in the
MSSM, although a fraction of the configurations are close to the detectability level.

\section{Antideuteron signal}
\begin{figure}[t]
\label{figure9}
\begin{center}
\includegraphics*[width=8.5cm]{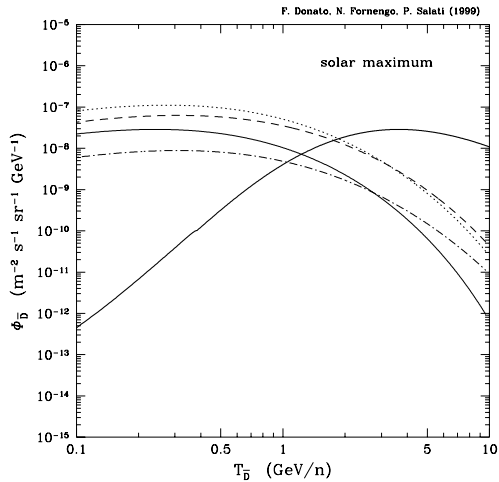}
\end{center}
\caption{Antideuteron flux vs. the antiproton kinetic energy \citep{noidbar}.
The solid bold line refers to the standard spallation component, while the other
lines show predictions for annihilation of WIMPs of different masses.}
\end{figure}
\begin{figure}[t]
\label{figure10}
\begin{center}
\includegraphics*[width=8.5cm]{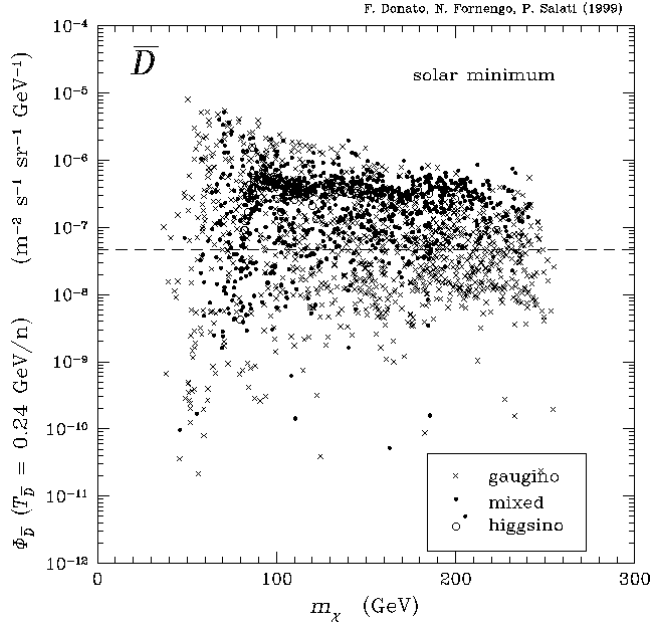}
\end{center}
\caption{Antideuteron flux at kinetic energy $T=0.24$ GeV produced by neutralino annihilation in the
galactic halo \citep{noidbar}. The supersymmetric model is a low--energy minimal supersymmetric standard model.
The horizontal line shows the predicted sensitivity of AMS for a 3 years data taking on board of the International
Space station.}
\end{figure}

The WIMP annihilation process in the galactic halo may produce also antideuterons, which then suffer analogous
diffusive and energy loss processes as the antiprotons. In \cite{noidbar} it has first been shown that the low
energy tail of the antideuteron flux offers a very good signal--to--background ratio, mostly due to kinematical
reasons. The low--energy antideuteron flux therefore offer a very good handle to detect a signal, contrary to antiprotons which currently seems better suited to set limits. Fig. 9 shows the antideuteron background and signals
vs. kinetic energies: below 1--3 GeV of kinetic energy the background is very much depressed, contrary to the
signal.

Fig. 10 then shows the predictions in the MSSM for one low--energy bin, compared to the expected sensitivity
for positive detection for the AMS detector on board of the ISS in a 3 years data taking. We see that a large
fraction of the configurations are accessible to detection. The proposed experiment GAPS \citep{gaps} is designed to access an order of magnitude more in sensitivity, which will allow to cover most of the configurations of the MSSM.

\clearpage

\section{Positron signal}
\begin{figure}[t]
\label{figure11}
\begin{center}
\includegraphics*[width=6.5cm]{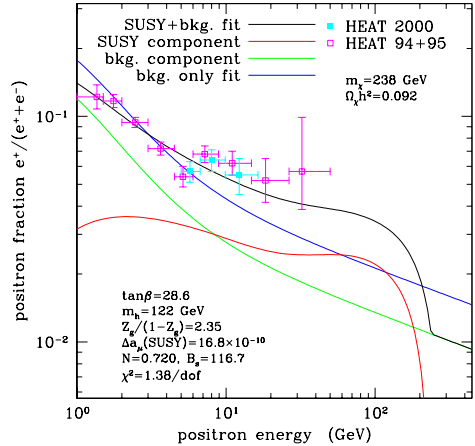}
\end{center}
\caption{Positron fraction vs. the positron energy. The data points are from HEAT.
The solid lines refer to background estimates and neutralino annihilation production
in the galactic halo (figure from \cite{positron}).}
\end{figure}

Dark matter annihilation in the galactic halo may also produce a positron signal. Fig. 11
shows the experimental result of the HEAT detector, which seems to indicate a bump
in the spectrum around energies of 10--30 GeV. This feature is hardly compatible with
the background estimates and this could point toward the presence of an exotic component.
Fig. 11 shows an example of positron production from neutralino annihilation. Even in this case
the spectral shape is difficult to reproduce, even though the agreement between theoretical
predictions and data improves. One has to notice also that the HEAT data would require a
sizeable overdensity of dark matter in order to explain the positron excess. This overdensity
should also be quite local in the Galaxy, since positron do not travel long distances in the
diffusive halo.

\section{Gamma rays signal}
\begin{figure}[t]
\label{figure12}
\begin{center}
\includegraphics*[width=8cm]{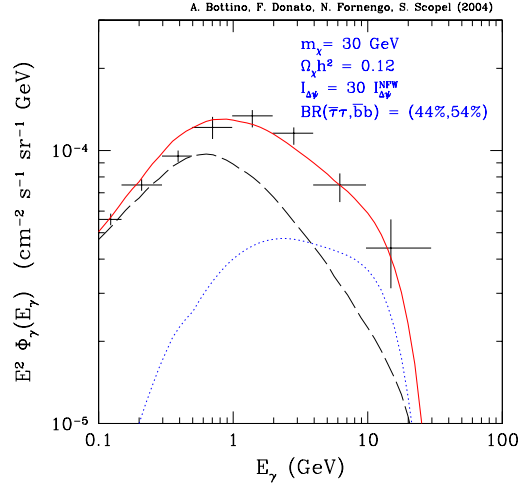}
\end{center}
\caption{Gamma ray flux vs. energy \citep{noiindirect}. The data points refer to EGRET.
The dashed line is an estimate of the gamma ray diffuse background, while the dotted line refers to
gamma rays production from a 30 GeV netralino annihilating in the galactic halo. The solid line is the total
gamma ray flux.}
\end{figure}
\begin{figure}[t]
\label{figure13}
\begin{center}
\includegraphics*[width=8cm]{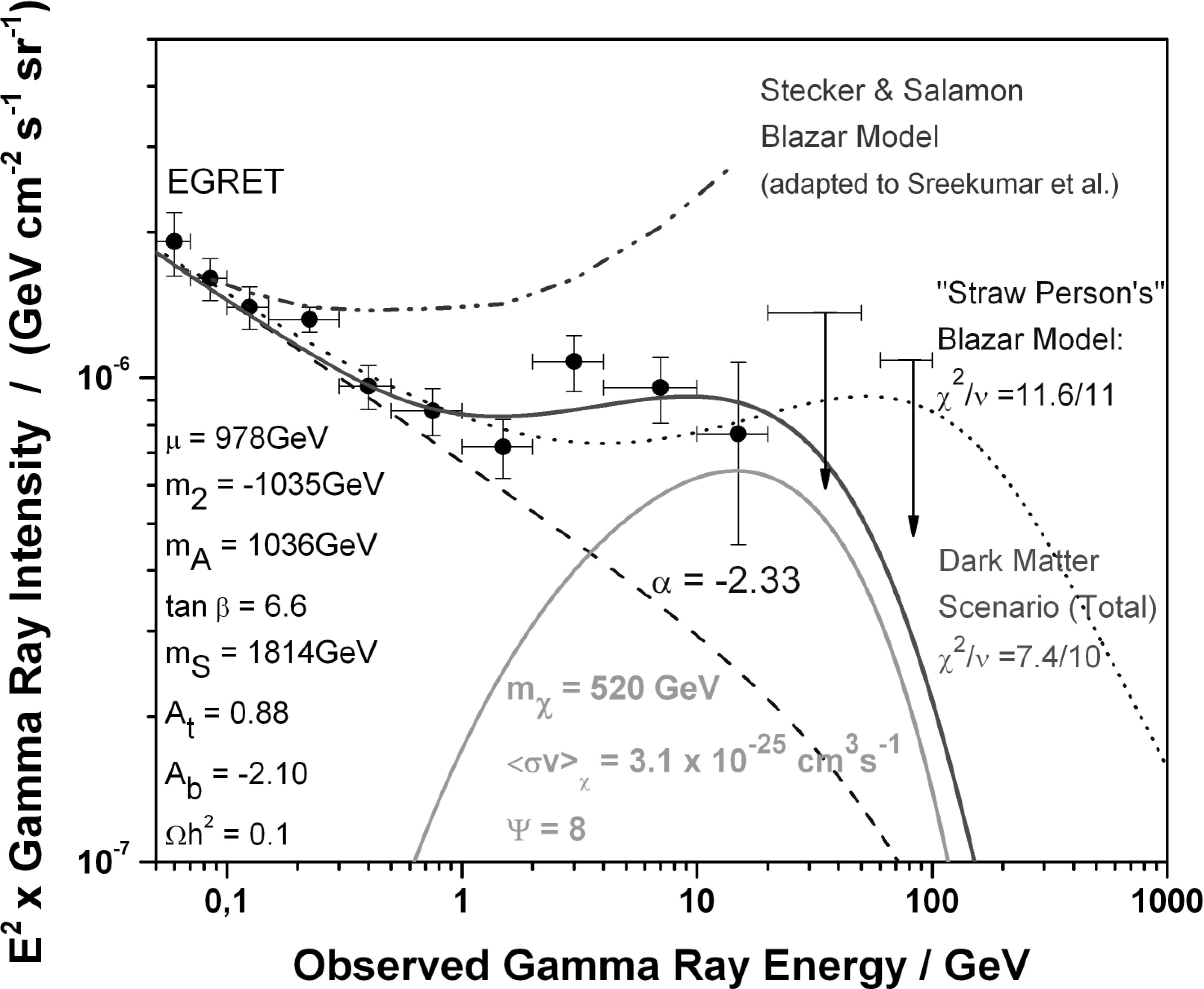}
\end{center}
\caption{Gamma ray flux vs. energy. The data points refer to EGRET. The dashed line is the standard bacgkround component. The lower solid line refer to a gamma ray contribution from a 520 GeV neutralino annihilation and
the upper solid curve is the sum of this contribution and the standard background.
The dot--dashed and dotted lines show the prediction of gamma rays produced in blazar models
(figure from \cite{mannheim}).}
\end{figure}
\begin{figure}[t]
\label{figure14}
\begin{center}
\includegraphics*[width=8cm]{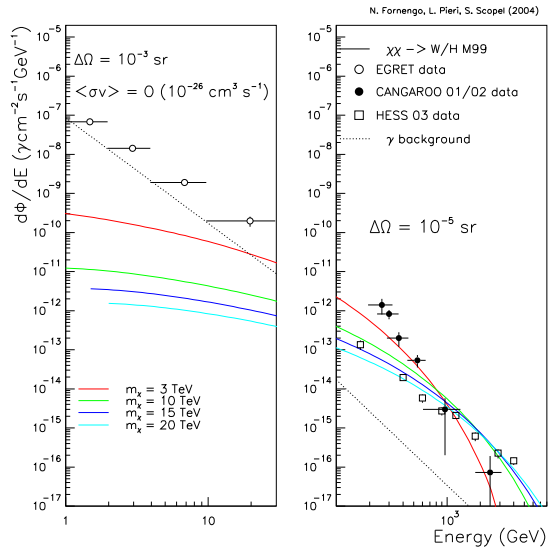}
\end{center}
\caption{Gamma ray flux vs. energy \citep{noilidia}. The right panel shows the comparison
among the HESS data and predictions for gamma ray flux produced by annihilation of TeV--scale neutralinos.}
\end{figure}
\begin{figure}[t]
\label{figure15}
\begin{center}
\includegraphics*[width=7.5cm]{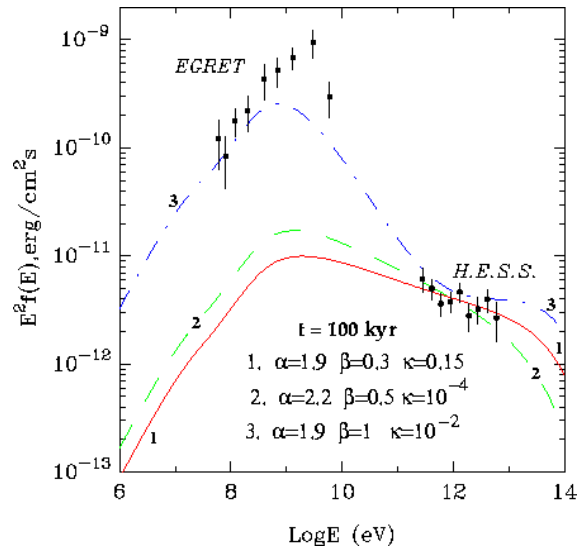}
\end{center}
\caption{Gamma ray flux vs. energy. The data points refer to EGRET and
HESS. The curves are predictions for astrophysical gamma--ray production in different galactic models
 (figure from \cite{aharonian}).}
\end{figure}
\begin{figure}[t]
\label{figure16}
\begin{center}
\includegraphics*[width=7cm]{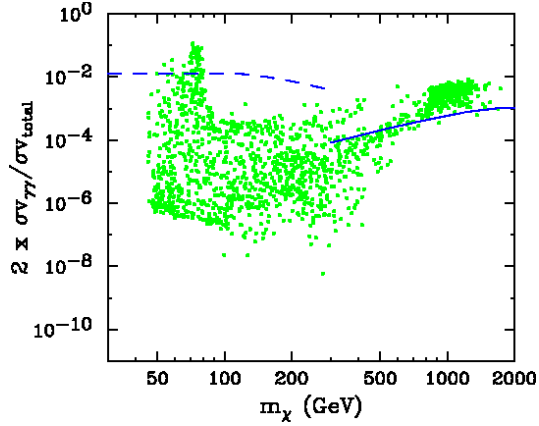}
\end{center}
\caption{Neutralino annihilation cross section into a gamma--ray line relative to the total annihilation cross section
vs. the neutralino mass. The dashes lines refer to the future expected sensitivity of GLAST and HESS
(figure from \cite{hooper}).}
\end{figure} 
\begin{figure}[t]
\label{figure17}
\begin{center}
\includegraphics*[width=7cm]{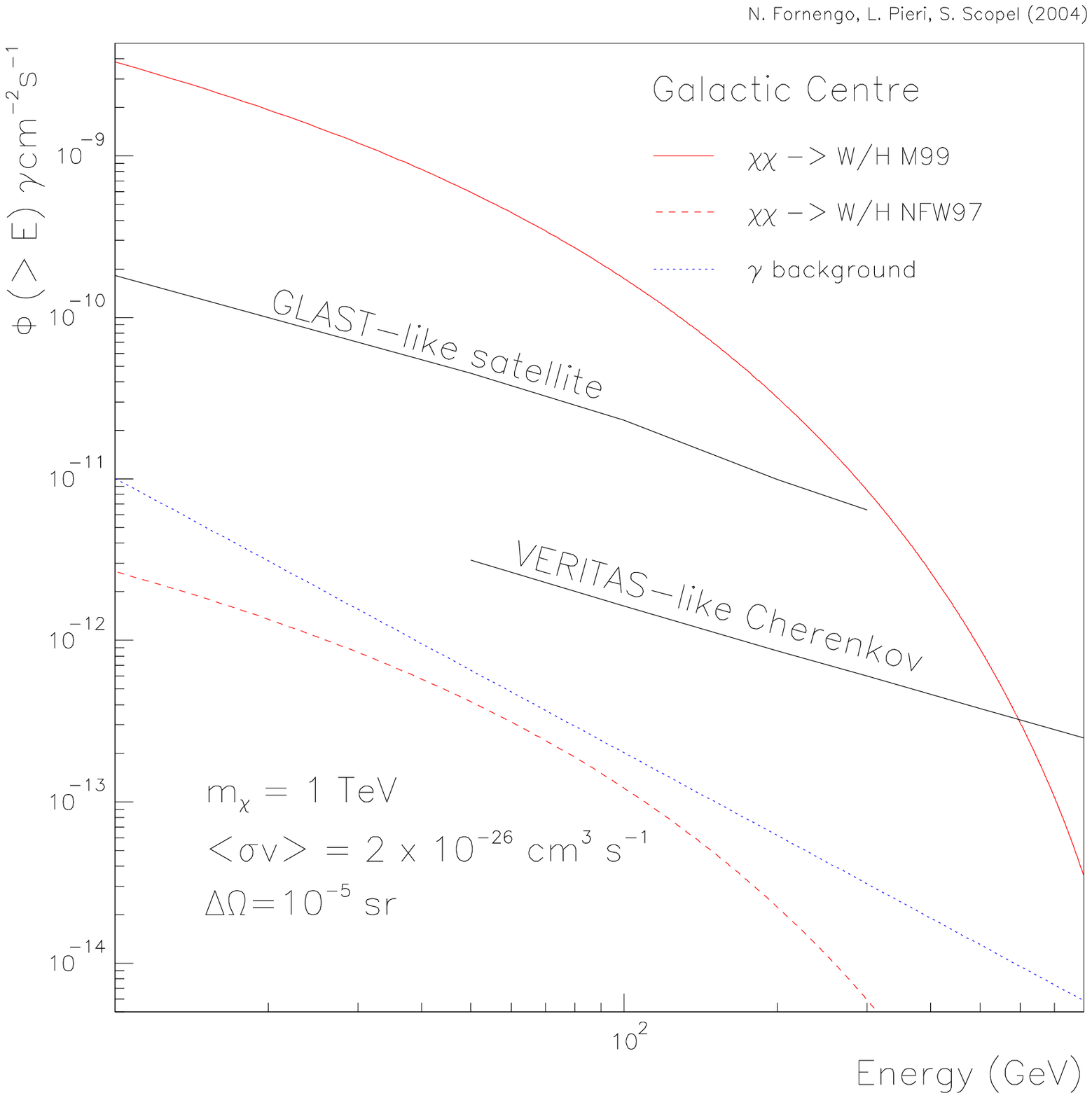}
\end{center}
\caption{Integral gamma--ray flux from M31 galaxy vs. the neutralino mass for a 1 TeV neutralino in the low--energy
minimal supersymmetric standard model\citep{noilidia}. The expected sensitivities of GLAST and VERITAS are shown.}
\end{figure}

Dark matter annihilation into gamma--rays is one of the best possibilites to be looked for. In this case
there is no uncertainty coming from propagation of the signal, contrary to the antimatter case. However
what are quite uncertain in this case are the properties of the source, expecially if one lookes toward
the galactic center which is the place where the dark matter density is larger (except in the case of the
presence of sizeable clumps, which nevertheless represent an additional element of uncertainty). 

At energies around 1--10 GeV the EGRET detector observes a possibile excess of gamma--rays over the
standard background. Fig. 12 shows these data together with a possibile explanation of the effect as
due to relatively light neutralinos in the MSSM. In this case, an overdensity factor of the dark matter
of the order of 30 above the case of a NFW density shape is needed. This is a typical situation for 
the gamma--rays case: typically overdensity factors are required in order to produce sizeable signals. 
These factors may be due to clumps along the line of sight toward the galactic center, or to steeper
density profiles, as predicted in some dynamical models of structure formation.

Fig. 13 shows that the EGRET excess could instead be explained in terms of standard astrophysical 
processes, with some (relatively minor) modifications of the standard background estimate (for
details, see for instance \cite{mannheim}). 

Another possible excess from the galactic center over the background could be present at larger energies, as seen, among others,  by the HESS detector. Fig. 14 shows this excess around the TeV energy scale, together with an interpretation in terms of neutralino annihilation. Although it may be possibile to explain this gamma ray production
in terms of DM contributions, nevertheless this flux is likely to be due to standard sources, as shown for 
instance in Fig. 15 and explained in \cite{aharonian}. The fact that the HESS data are very well reproduced by
a power low profile, this points toward the interpretation in terms of a standard source. In this case, 
interesting limits on the neutralino annihilation component could be set \citep{hooper}.

Neutralinos may also produce a gamma--ray line, in addition to a diffuse flux. Since this is a 1--loop
process, it is suppressed as compared to the diffuse one. The prospects of detecting such a line are
not very good, although GLAST and HESS could access some configurations, as shown in Fig. 16.

A possibility of looking for gamma--rays from dark matter annihilation in external galaxies has also been
proposed. Fig. 17 shows some prospects for searching for this contributions from M31 galaxy with next
generation detectors.

\section{Neutrinos from Earth and Sun}
\begin{figure}[t]
\label{figure18}
\begin{center}
\includegraphics*[width=8cm]{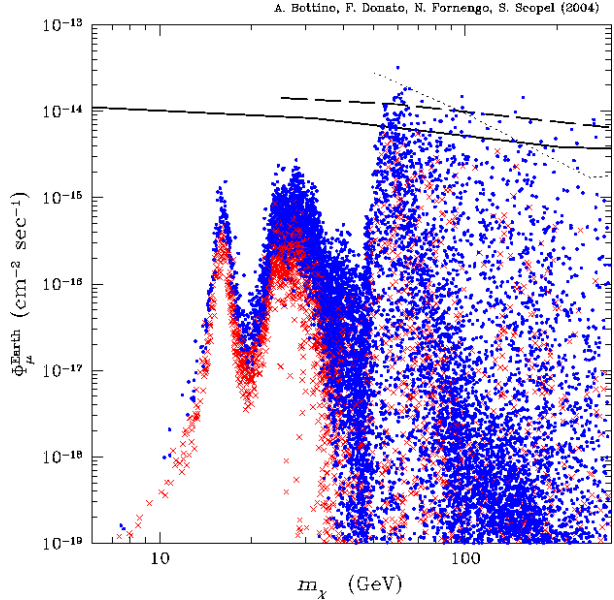}
\end{center}
\caption{Upgoing muon flux from neutrinos produced by neutralino annihilation in the center of the Earth, 
in the low--energy minimal supersymmetric standard model \citep{noiindirect}. The solid/dashed/dotted
lines refer to upper limits from SuperKamiokande/MACRO/Amanda experiments.}
\end{figure}
\begin{figure}[t]
\label{figure19}
\begin{center}
\includegraphics*[width=8cm]{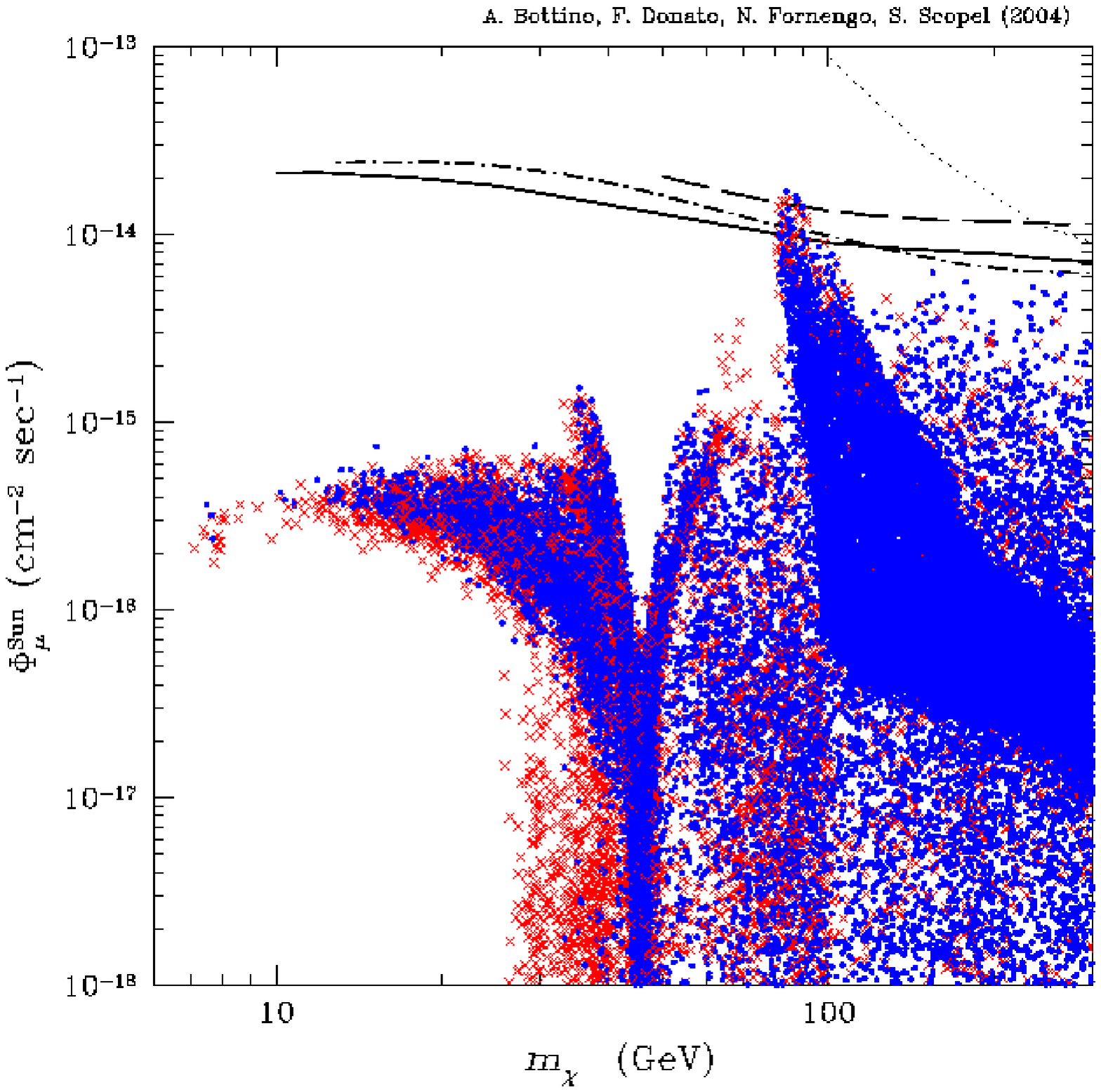}
\end{center}
\caption{Upgoing muon flux from neutrinos produced by neutralino annihilation in the Sun, 
in the low--energy minimal supersymmetric standard model \citep{noiindirect}. The solid/dot--dashed/dashed/dotted
lines refer to upper limits from SuperKamiokande/Baksan/MACRO/Amanda experiments.}
\end{figure}
\begin{figure}[t]
\label{figure20}
\begin{center}
\includegraphics*[width=7cm]{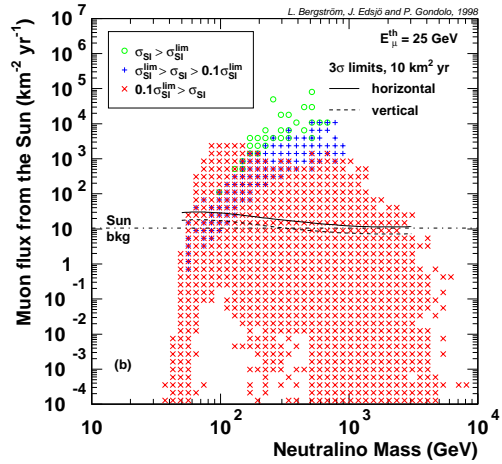}
\end{center}
\caption{Predictions of upgoing muon flux from neutrinos produced by neutralino annihilation in the Sun for a large--area neutrino telescope, in the low--energy minimal supersymmetric standard model. The solid and dashed lines
show the expected sensitivities. The dot--dashed lines show the neutrino production from cosmic--rays interactions
with the solar corona (figure from \cite{bergstrom}).}
\end{figure}

Finally, we have the possibility to look for a neutrino signal produced in the center of
the Earth or the Sun where DM may accumulate after gravitational capture. The muon neutrino component
of this flux is then detected as upgoing muons in a neutrino telescope. Fig. 18 and Fig. 19 show the
theoretical predictions in the MSSM for the upgoing muon flux coming from the Earth and the Sun, compared to the
current experimental sensitivities. We see that the neutrino signal is not suited for light neutralinos,
since in this case the flux is too soft. Larger mass neutralinos may easily produce harder neutrinos
which then can produce muons above the detector threshold.

Fig. 20 shows prospects for a km--size detector, for which the threshold energy is larger and therefore
only large mass neutralinos could be accessed.


\begin{thebibliography}{}


\bibitem[Bernabei et al.(2003)]{dama} Bernabei, R., Belli, P. Cappella, T., et al., 
Dark Matter search, 
2003, Riv. N. Cim. 26 n. 1, 1--74.

\bibitem[Bottino et al.(2005a)]{noidirect} Bottino, A., Donato, F., Fornengo, N., \& Scopel, S.,
Do current WIMP direct measurements constrain light relic neutralinos?,
2005, Phys. Rev. D72, 083521.

\bibitem[Bottino et al.(2003)]{noilow} Bottino, A., Donato, F., Fornengo, N., \& Scopel, 
Lower Bound on the Neutralino Mass from New Data on CMB and Implications for Relic Neutralinos,
2003, Phys. Rev. D68, 043506.

\bibitem[Akerib et al.(2005)]{cdms} Akerib, D.S., Armel--Funkhouser, M. S., Attisha, M.J., et al.,
Exclusion limits on the WIMP-nucleon cross section from the first run of the Cryogenic Dark Matter 
Search in the Soudan Underground Laboratory,
2005, Phys. Rev. D 72, 052009.

\bibitem[Baek et al.(2005)]{munoz} S. Baek, D.G. Cerdeno, Y.G. Kim, P. Ko , C. Munoz,
Direct detection of neutralino dark matter in supergravity,
2005, JHEP 0506, 017.

\bibitem[Donato et al.(2004)]{noipbarsalati} Donato, F., Fornengo, N., Maurin, D., Salati, P., \& Taillet, R., 
Antiprotons in cosmic rays from neutralino annihilation,
2004, Phys. Rev. D69, 063501.

\bibitem[Orito et al.(2000)]{bess95} Orito, S., Maeno, T., Matsunaga, H., et al.
Precision Measurement of Cosmic-Ray Antiproton Spectrum,
2000, Phys. Rev. Lett. 84, 1078--1081.

\bibitem[Maeno et al.(2001)]{bess98} Maeno, T., Orito, S., Matsunaga, H., et al.,
Successive measurements of cosmic-ray antiproton spectrum in a positive phase of the solar cycle,
2001, Astropart. Phys. 16, 121--128.

\bibitem[Aguilar et al.(2002)]{ams} Aguilar, M., Alcaraz, J., Allaby, J., et al.,
The Alpha Magnetic Spectrometer (AMS) on the International Space Station: Part I Ð 
results from the test flight on the space shuttle,
2002, Phys. Rep. 366, 331--405.

\bibitem[Boezio et al.(2001)]{caprice} Boezio, M., Bonvicini, V., Schiavon, P., et al.
The Cosmic-Ray Antiproton Flux between 3 and 49 GeV,
2001, Astrophys. J. 561, 787--799.

\bibitem[Bottino et al.(2004)]{noiindirect} Bottino, A., Donato, F., Fornengo, N., \& Scopel, S., 
Indirect signals from light neutralinos in supersymmetric models without gaugino mass unification,
2004, Phys. Rev. D70, 015005.

\bibitem[Bottino et al.(2005b)]{noipbar} Bottino, A., Donato, F., Fornengo, N., \& Salati, P., 
Antiproton fluxes from light neutralinos,
2005, Phys. Rev. D72, 083518.

\bibitem[Donato et al.(2000)]{noidbar} Donato, F., Fornengo, N. \& Salati, P., 
Antideuterons as a Signature of Supersymmetric Dark Matter,
2000, Phys. Rev. D62, 043003.

\bibitem[Baltz et al.(2002)]{gaps} Mori, K., Hailey, C. J., Baltz, E. A., et al.,
A Novel Antimatter Detector Based on X-Ray Deexcitation of Exotic Atoms,
2002, Astrophys. J. 566, 604--616.

\bibitem[Baltz et al.(2002)]{positron} Baltz, A., Edsjo, J., Freese, K., \& Gondolo, P., 
Cosmic ray positron excess and neutralino dark matter,
2002, Phys. Rev. D 65, 063511.

\bibitem[Bergstrom et al.(1998)]{bergstrom} Bergstrom, L., Edsjo, J., \& Gondolo, P., 
Indirect detection of dark matter in km-size neutrino telescopes,
1998, Phys. Rev. D 58, 103519.

\bibitem[Elsaesser et al.(2005)]{mannheim} Elsaesser, D., \& Mannheim, K.,
Supersymmetric Dark Matter and the Extragalactic Gamma Ray Background,
2005, Phys. Rev. Lett. 94, 171302.

\bibitem[Fornengo et al.(2004)]{noilidia} Fornengo, N., Pieri, L., \& Scopel, S., 
Neutralino annihilation into gamma-rays in the Milky Way and in external galaxies,
2004, Phys. Rev. D70, 103529.

\bibitem[Aharonian et al.(2004)]{aharonian} Aharonian, F., \& Neronov, A.,
TeV gamma rays from the Galactic Center,
2005, astro--ph/0503354.

\bibitem[Zaharijas et al.(2004)]{hooper} Zaharijas, G., \& Hooper, D., 
Challenges in detecting gamma-rays from dark matter annihilations in the galactic center,
2006, Phys. Rev. D 7073, 103501.

\end{thebibliography}
\end{document}